\documentclass[12pt]{article}
\usepackage{graphicx}
\usepackage[square]{natbib}
\usepackage{amsmath,amssymb}
\usepackage{rotating}
\usepackage{multirow}
\usepackage[footnotesize]{caption}
\title{\Huge On the fish-like swimming \\ of linked bodies \\with and without 
 skin
\vskip 3cm }
\author{ \Large  {J. B. Kajtar \quad and \quad J. J. Monaghan }  \\
 { \small School of Mathematical Sciences } \\ {\small Monash University,  Vic 3800 Australia.}\\
{ \small email: joe.monaghan@sci.monash.edu.au }}\date {\small 17 December 2009}

\begin{document}

\maketitle
\vskip3cm

\newpage
\thispagestyle{empty}

\begin{abstract}  
In this paper we study the two dimensional motion of three linked rigid bodies moving through a fluid. The bodies change their orientation relative to each other in a way which mimics the swimming of fish. In contrast to previous simulations the bodies are connected by an elastic skin. The skin responds to the movement of the bodies and the pressure of the fluid and alters the flow around the bodies. In particular it prevents fluid moving between them. The system of bodies and skin is similar in appearance to a swimming leech or tadpole depending on the relative size of the bodies. We simulate the system using SPH, with three types of particles: liquid particles, boundary force particles determining the surface of the rigid bodies, and skin particles defining the elastic skin. The latter interact with each other, and with the boundary force particles to which they are anchored, by linear spring forces.  The boundary force particles and the skin particles interact with the fluid particles by pair forces which are similar to the forces used in the Immersed Boundary method. The algorithm is based on a Lagrangian, and the equations of motion conserve linear momentum exactly and angular momentum very accurately. We compare the motion of rigid bodies with and without skin keeping the total mass of the bodies plus skin fixed. When the ellipses are identical, and the forward gait is used, the bodies swim $\sim 13\%$ faster when they are connected by skin, and they require less energy. When the ellipses have different sizes, with the front ellipse largest, they travel $\sim 39\%$ faster and use less energy. In the case of the turning gait, the identical ellipses turn faster with skin and use less energy, but the different sized ellipses turn more slowly with skin. The algorithm is simple and robust and can be applied to bodies of arbitrary shape and in domains which include free surfaces and stratified fluids.

\end{abstract}
\newpage

\section{Introduction}

Most marine creatures swim by changes in body shape.  During these changes the outer surface remains smooth because of the elastic properties of the skin and tissue of the body. In earlier papers we  approximated the motion of marine creatures by considering linked rigid bodies moving in two dimensions in response to changes in the angles between them, but neglected the effects of skin and tissue (Kajtar and Monaghan 2008). The bodies we considered were three identical ellipses connected by virtual rods which allowed fluid to move between the bodies.  Our high Reynolds number results were in good agreement with the two dimensional inviscid calculations of Kanso et al., (2005) and Melli et al., (2006) and for lower Reynolds numbers they were in good agreement with  the viscous calculations  of Eldredge (2006, 2007, 2008).  

Although this model of swimming creatures is very crude it gives a surprisingly accurate prediction of the motion of a leech (Kajtar and Monaghan, 2010).  Nevertheless it is desirable to improve the model to bring it closer to the motion of actual marine creatures.  In particular we wish to mimic the elastic properties of their  bodies while eliminating the flow between them.  A simple way to do this is to connect the bodies by an elastic skin.  We do not claim that this is anything but a crude representation of actual tissue, but it represents important features of such tissue and opens the way to represent it more accurately.  We note, in particular, that an elastic surface will deform under pressure forces from the liquid. 

We simulate the system using SPH, with separate  particles for the liquid, the boundary of the rigid bodies,  and the skin, and we derive the inviscid equations from a Lagrangian variational principle. The viscous equations then follow by adding a standard SPH viscous term.  We apply this algorithm to both straight line motion and to turning motion.  The algorithm conserves linear momentum to within round-off error.  The time stepping introduces relative errors in the conservation of angular momentum which are typically $10^{-6}$. However, because we approximate the infinite fluid by a periodic domain, there is a larger change in the angular momentum  because periodic boundaries do not conserve the angular momentum of a particle system.  An alternative approach to the simulation of swimming fish is the method described by Borazjani et al. (2008) and Borazjani and Sotiropoulos (2008, 2009, 2010) who use an Immersed Boundary method (Peskin, 1977, 2002).  In their method the fish body is triangulated and treated as an immersed boundary which moves  in a specified way. This method is similar to our method except that we use particles for the entire system while they use particles only to specify the fish-body surface which is not allowed to deform under liquid pressure forces.  Furthermore, their algorithm is currently designed for straight line motion so that the effect of torques on the body are not included. 

The plan of this paper is to discuss the  SPH equations of motion and the modeling of the skin.  We then compare the speed and power output  of three identical linked ellipses with and without the skin both forward and for turning motions.  Finally we apply the method to a system of three different ellipses linked as before. It is trivial to apply the algorithm to the swimming of bodies through free surfaces, and to studies of swimming in stratified media.

\section{SPH equations for the fluid}
\setcounter{equation}{0}
The continuum equations we solve are the Navier-Stokes equations with boundaries formed by parts of rigid bodies and sections of skin. Apart from the introduction of the skin, the equations are the same as those we have simulated before (Kajtar and Monaghan, 2008). To simplify the paper  we give the details of the SPH equations and refer the reader to the continuum equations described by Kajtar and Monaghan (2008). 
  
\subsection{The acceleration equations}
  
In the following we use $a$ and $b$ for the labels of the liquid SPH particles, $j$ for the label of boundary force particles on the rigid bodies, and $\sigma$ as the label for the skin particles. We write the equation of motion for the liquid particle $a$ in the form
\begin{equation}
\frac{d {\bf v}_a}{dt} = {\bf F}_a({\it fluid}) + {\bf F}_a({\it body}) + {\bf F}_a({\it skin}),
\end{equation}
where
\begin{equation}
 {\bf F}_a({\it fluid})  = - \sum_b m_b \left (  \frac{P_a}{\rho_a^2 }  +\frac{P_b}{\rho_b^2} + \Pi_{ab}    \right) \nabla_a W_{ab},
\end{equation}

\begin{equation}
 {\bf F}_a({\it body})  = - \sum_j m_j \left (  \frac{P_a}{\rho_a^2 }  +\frac{P_j}{\rho_j^2} + \Pi_{aj}    \right) \nabla_a W_{aj} + \sum_{k=1}^{N_b} \sum_{j \in B(k) }m_j {\bf r}_{aj}f(|{\bf r}_{aj}| ),
\end{equation}
and 
\begin{equation}
 {\bf F}_a({\it skin})  = - \sum_\sigma m_j \left (  \frac{P_a}{\rho_a^2 }  +\frac{P_\sigma}{\rho_\sigma^2} + \Pi_{a\sigma}    \right) \nabla_a W_{a\sigma} + \sum_{k=1}^{N_s} \sum_{\sigma \in S(k) }m_\sigma {\bf r}_{a \sigma} f(| {\bf r}_{a\sigma} | ).
\end{equation}

$ {\bf F}_a({\it fluid})$ is the pressure and viscous force per unit mass due to the other fluid particles.   ${\bf F}_a({\it body}) $ is the force per unit mass due to the rigid bodies.  It consists of two parts. The first is a direct pressure interaction which is a  result of deriving the equations of motion from a variational principle using the continuity equation as a constraint. The second is based on Sirovich's formulation of the effects of boundaries (Sirovich, 1967, 1968) which we have discussed elsewhere (Monaghan and Kajtar, 2009). Our prescription for this is similar to  the Immersed Boundary method.  A typical boundary particle $j$ on the surface of  the rigid body exerts a repulsive force $m_a m_j {\bf r}_{aj} f( | {\bf r}_{aj} | )$ on fluid particle $a$ along the line joining their centers. Here and elsewhere ${\bf r}_{aj} = {\bf r}_a - {\bf r}_j$.  Correspondingly, fluid particle $a$ exerts an equal but opposite force $m_j m_a {\bf r}_{ja} f( | {\bf r}_{aj} | )$ The form of the function $f( | {\bf r}_{aj} | ) $  is chosen so that it mimics a delta function and provides a force on the fluid particle which is normal to the surface of the body to a very close approximation (Monaghan and Kajtar 2009). The force per unit mass  due to the skin particles ${\bf F}_a({\it skin})$ is identical except the summations are over skin particles.

In these equations $m_b$ is the mass of particle $b$, $P_b$  and $\rho_b$ are the pressure and density at the position ${\bf r}_b$ of particle $b$. We use the same equation of state to determine $P$ in terms of $\rho$ as that used by Kajtar and Monaghan (2008). Further details are given in \S\ref{sec:eqnstate}. This equation of state makes the fluid weakly compressible.  $\Pi_{ab}$ specifies the viscous interaction between particles $a$ and $b$.  We use the same form of the viscous interaction as in Kajtar and Monaghan (2008).   $W_{ab}$ denotes the smoothing kernel $W({\bf r}_a - {\bf r}_b, \bar h_{ab})$ and $\nabla_a$ denotes the gradient taken with respect to the coordinates of particle $a$. In this paper $W$ is the fourth degree Wendland kernel for two dimensions (Wendland, 1995), and has support $2\bar h_{ab}$.  In the present calculations the $\bar h_{ab}$ used in $W_{ab}$ is an average $\bar h_{ab}= (h_a + h_b)/2$.  The choice of $h$ is discussed in detail by Monaghan (1992, 2005). In this paper we choose the initial $h$ for any particle to be 1.5 times the initial particle spacing but, thereafter, it is determined by the local density.  The total number of bodies is $N_b$ and the total number of skin segments is $N_s$.  $B(k)$ denotes the set of labels associated with body labelled $k$ and $S(k)$ denotes the set of labels associated with skin segment $k$.

The acceleration of the center of mass ${\bf R}_k$ of  body $k$ with mass $M_k$ takes the form
\begin{equation}
M_k \frac{ d {\bf V}_k}{dt} = \sum_{j \in S_k} {\bf f}_j +   {\bf F}_k,
\end{equation}
where ${\bf F}_k$ is a constraint force associated with the specification of the angles $\varphi$ between the bodies. The torque equation is 
\begin{equation}
I_k \frac{ d \Omega_k}{dt} = \sum_{j \in S_k} ({\bf r}_j - {\bf R}_k) \times{\bf f}_j + \tau_k,
\end{equation}
where $\tau_k$ is the torque associated with the constraints. The constraint forces and torques are discussed further in \S\ref{sec:constraints}.

The force ${\bf f}_j$ on boundary particle $j$ of body $k$ is given by 
\begin{equation}
  {\bf f}_j  = - m_j \sum_a m_a \left (  \frac{P_a}{\rho_a^2 }  +\frac{P_j}{\rho_j^2} + \Pi_{aj}  \right) \nabla_j W_{aj} + \sum_a m_j m_a {\bf r}_{ja} f(|{\bf r}_{ja}  | )  + {\bf f}_j(\it skin),
\end{equation}
where the first term is the force on the body particle $j$ due to the pressure and viscous stress of the fluid, the second term is the reaction force  arising from the forces on the fluid due to the second term in (2.3). The third term is the skin force which we discuss in \S\ref{sec:skin}.

The acceleration of skin particle $\sigma$ is due to a pressure interaction with the fluid, a repulsive force interaction with the fluid (these are similar to those discussed for rigid body boundary particles), and a force per unit mass due to neighbouring skin and/or body particles,
\begin{equation}
\frac{d^2{\bf r}_\sigma}{dt^2} = - \sum_\sigma m_a \left (  \frac{P_a}{\rho_a^2 }  +\frac{P_\sigma}{\rho_\sigma^2} + \Pi_{a\sigma}    \right) \nabla_\sigma W_{a\sigma} + \sum_a m_a {\bf r}_{\sigma a} f( |{\bf r}_{\sigma a} |  ) + {\bf f}_\sigma ({\it body}) + {\bf f}_\sigma ({\it skin}).
\end{equation}
The third term is the interaction with the bodies to which the skin is anchored ($ \S\ 2.2 )$.  The interaction between any pair of SPH particles is along the line of centres, and the force on one particle is opposite to the force on the other. As a consequence, linear and angular momentum are conserved.

The force function between the skin and either the boundary or skin particles has the following form:
\begin{equation}
f(|{\bf r}_{bj}|) = \begin{cases}
\frac{ BW(|{\bf r}_{bj}|)}{r^2}   & \text{;  $r_{bj} \le 2h$}, \\                     
0 & \text{; $r_{bj} > 2h$},
\end{cases}
\end{equation}
where $B = V^2/{\bar m}$, $\bar m$ is the mass of a liquid particle, and  $V$ is the maximum speed of the fluid. $V$ is estimated at the beginning of the calculation and thereafter held constant. $W$ is the Wendland cubic kernel normalized to 1 at the center. With $ q = r/h$ it has the form
\begin{equation}
W(r) = \begin{cases}
\frac18(1+3q/2)(2-q)^3  & \text{;  $r \le 2h$}, \\                     
0 & \text{; $r > 2h$},
\end{cases}
\end{equation}

\subsection{Skin}\label{sec:skin}

In the present two dimensional study  a section of the  elastic skin is a single line of skin particles connected by spring-like forces. Only neighboring particles interact.  When the skin is stretched to length $L$, the skin tension is $T = \kappa L$, where $\kappa$ is a constant spring force per unit length. The skin thickness is $\ell_s$, and the mass of each skin particle $\sigma$ is $m = \rho_\sigma \ell_s \Delta_s$, where $\rho_\sigma$ denotes the skin density (which is constant), and $\Delta_s$ the initial skin particle spacing.  

\begin{figure}[htbp]
\begin{center}
\includegraphics[width=0.5\textwidth]{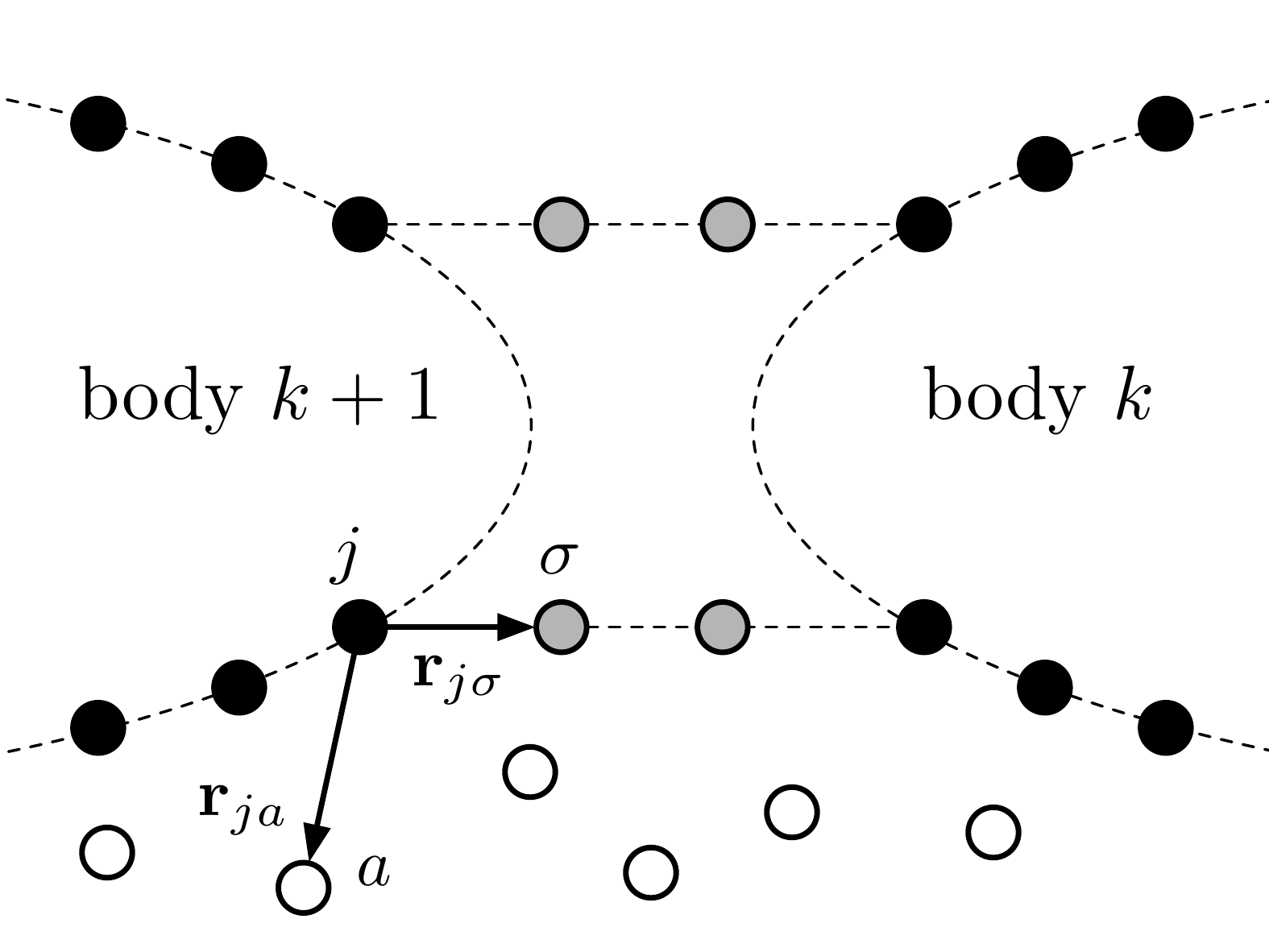}
\caption{An illustration of the placement of body and skin particles. Filled circles represent body particles and are placed around the perimeter of the elliptical bodies. Shaded circles represent skin particles. In this case, the skin is taut, but in general it flexes under the pressure of the fluid. The open circles represent fluid particles. The following vectors along which the forces are calculated have also been indicated: on fluid particle $a$ due to body particle $j$, and on skin particle $\sigma$ due to $j$.}
\label{fig:bodyandskin}
\end{center}
\end{figure}

Skin particles interact with the fluid with the same boundary force as body particles. Skin particles can move in response to the forces acting on them, whereas body particles only move when the body to which they are attached moves. A typical configuration of body, skin and liquid particles is illustrated in Figure \ref{fig:bodyandskin}. 

The elastic force on a skin particle with  label $\sigma$ due to a neighbouring skin particle with label $\sigma'$,  is given by
\begin{equation}\label{eqn:skinforce}
{\bf f}_\sigma = \kappa ({\bf r}_{\sigma'} - {\bf r}_{\sigma}).
\end{equation}
The skin particles are always labelled such that $\sigma+1$, $\sigma$ and $\sigma-1$ are contiguous. In the case where the neighbouring particle is a body particle (to which the skin is attached), the label $\sigma'$ is replaced by that of the body particle.

The continuum limit of our skin shows that the speed of a transverse wave propagating along the skin is 
\begin{equation}
v_s = \sqrt{\frac{T \Delta_s}{m}}.
\end{equation}
The parameter $\kappa = T/ \Delta_s$ is then given by
\begin{equation}
\kappa = \frac{v_s^2 m}{\Delta_s^2}.
\end{equation}
We choose the skin parameters so that $v_s$ is comparable to the speed of sound $c_s$ of our slightly compressible fluid to ensure the CFL condition from both speeds is similar. The details are discussed in connection with time stepping.

The form of body particle-skin interaction ${\bf f}_j (\it skin)$ in (2.7) is determined by the fact that only one of the boundary particles on a given rigid body can connect with a specified section of skin. The first and last particles on each section connect with a boundary particle of a body. In general, for this two dimensional problem, a body has four such connecting boundary particles, while the first and last bodies have two connecting particles. For any given boundary particle of a rigid body it either connects to a skin section or it doesn't. If it does connect it does so by an elastic force term. Thus, for (2.7)
\begin{equation}
{\bf f}_j (\it skin) = 
\begin{cases}
\kappa({\bf r}_\sigma - {\bf r}_j)   & \text {; $j$ connected to $\sigma$}, \\                     
0, & \text{; $j$ not connected to $\sigma$}.
\end{cases}
\end{equation}

\subsection{The constraints}\label{sec:constraints}

\begin{figure}[htbp]
\begin{center}
\includegraphics[width= 0.9 \textwidth]{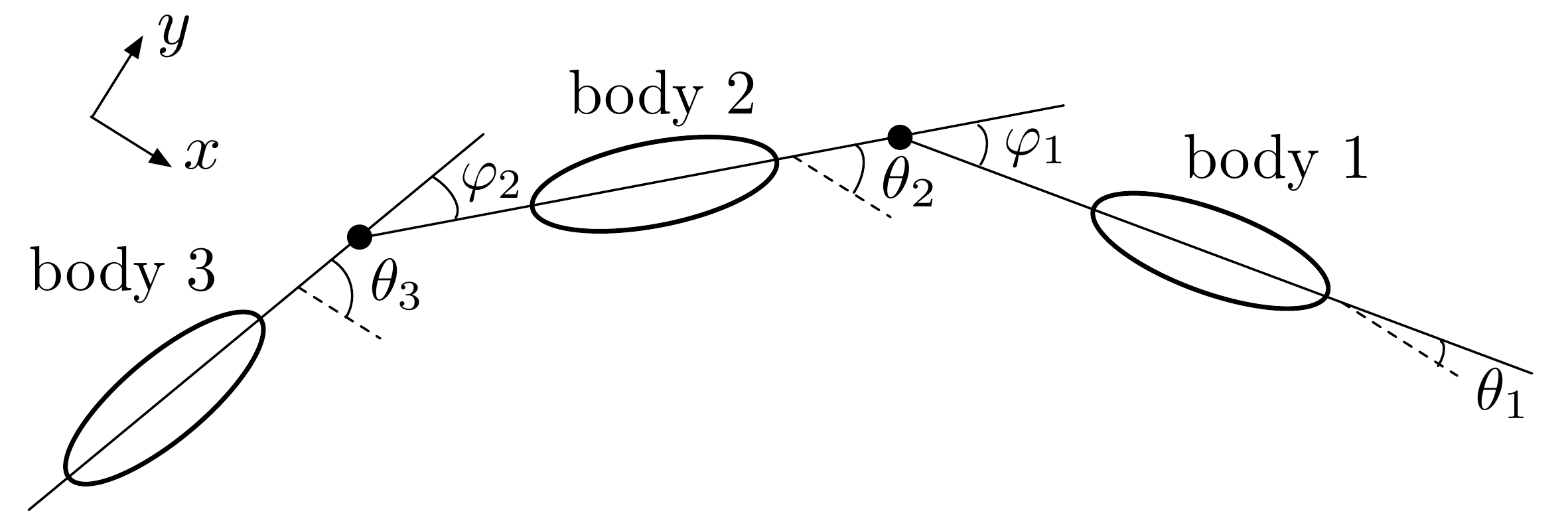}
\caption{The configuration of the bodies (assumed to be ellipses). The skin is not shown. The link position is denoted by a filled circle. The straight line through a body passes through its centre of  mass and is assumed to be rigidly attached to the body with one end attached to the link.  The angles $\theta$ are defined relative to a fixed direction in space (shown by parallel dotted lines) which is taken to be the $x$ axis of a cartesian coordinate system in our calculations.  The angles $\varphi$ determine the gait and are specified functions of  time.}
\label{fig:bodyangles}
\end{center}
\end{figure}

The angle $\theta_k$ which fixes the rotation of body $k$ is defined as the positive rotation of a line fixed in the body from the $x$ axis of a cartesian coordinate system fixed in space.  For simplicity we assume the line fixed in the body is an axis of symmetry. The constraint conditions on the angles are 
\begin{equation}
\varphi_m = \theta_{m+1} - \theta_m,
\end{equation}
where $m$ is the link number and $\varphi_m$ is a specified function.  The form of the $\varphi_m$ determines the gait of the bodies.  For the examples we consider here there are three bodies and two links as shown in Figure \ref{fig:bodyangles} with the skin removed for clarity.  In the simplest case $\varphi_m$ is a function of $t$ but, in general, it depends on other variables.  For example, in a biological problem, it could depend on the centre of mass coordinates in such a way that the fish slows down when it enters a region where food is abundant.  

In addition to the constraints on the angles there are constraints associated with the links.  We assume the link, or pivot, is  at a distance $\ell_k$ from the centre of mass of body $k$.  The condition on the $X$ components of the centres of mass of bodies $k$ and $k+1$ is that the $X$ coordinate of the link between them is given by
\begin{equation}\label{eqn:xcons}
X_k  - \ell_k  \cos{(\theta_k)}  -X_{k+1} -  \ell_{k+1} \cos{(\theta_{k+1})} = 0.
\end{equation}
Similarly, the $Y$ constraint is 
\begin{equation}\label{eqn:ycons}
Y_k  - \ell_k  \sin{(\theta_k)}  -Y_{k+1} - \ell_{k+1} \sin{(\theta_{k+1})} = 0.
\end{equation}
These constraints enable the coordinates of the centers of mass of the bodies, and their angles $\theta$ to be written in terms of those of any selected body.  Similarly, by differentiating the constraint conditions with respect to time, the velocities $\dot X$ and $\dot Y$ and angular velocity $\Omega$ of the bodies can  be written as functions of the same selected body.  The number of degrees of freedom (coordinates and velocities) of $N$ linked bodies in two dimensions is therefore 6 compared with the $6N$ degrees of freedom of $N$ independent bodies in two dimensions.  If the $\varphi_m$ are functions of $t$ alone it is possible to reduce the equations of motion to those involving the coordinates and velocities of one of the bodies.  This can also be done when the $\varphi_m$ are functions of both coordinates and time but it is inconvenient to eliminate variables and, in our view, simpler to take account of the constraints by using Lagrange multipliers. For that reason we use Lagrange multipliers even though, in the applications to be described in this paper, the $\varphi_m$ are functions of $t$ only.  For the case of three bodies we have two links and therefore 6 constraints. 
 
We denote the Lagrange multipliers for the $X$, $Y$ and $\theta$ constraints of link $m$ by  $\lambda_X^{(m)}$, $\lambda_Y^{(m)}$ and $\lambda_\theta^{(m)}$ respectively.  Using standard methods  for holonomic constraints (e.g. Landau and Lifshitz, 1976) we find the following expressions for the constraint forces ${\bf F}_k$ and torques  $ \tau_k$ for the various bodies. For bodies 1, 2 and 3 respectively,
\begin{align}
{\bf F}_1 &= (\lambda_X^{(1)} , \lambda_Y^{(1)}), \\
{\bf F}_2 &= (-\lambda_X^{(1)} , -\lambda_Y^{(1)}) + (\lambda_X^{(2)},\lambda_Y^{(2)}), \\
{\bf F}_3 &= (-\lambda_X^{(2)},-\lambda_Y^{(2)}).
\end{align}
These constraint forces do not affect the total linear momentum of the bodies because they sum to zero. 

The constraint torques on bodies 1, 2 and 3 respectively are 
\begin{align}
\tau_1 &= -\lambda_{\theta}^{(1)}+ \lambda_X^{(1)} \ell_1 \sin{(\theta_1)} - \lambda_Y^{(1)} \ell_1 \cos{(\theta_1)}, \\
\tau_2 &= \lambda_{\theta}^{(1)} - \lambda_{\theta}^{(2)}+ \lambda_X^{(1)} \ell_2 \sin{(\theta_2)}  + \lambda_X^{(2)} \ell_2  \sin{(\theta_2)}- (\lambda_Y^{(1)}+ \lambda_Y^{(2)}) \ell_2 \cos{(\theta_2)}, \\
\tau_3 &= \lambda_{\theta}^2+ \lambda_X^{(2)} \ell_3 \sin{(\theta_3)} - \lambda_Y^{(2)} \ell_3 \cos{(\theta_3)}.
\end{align}

The Lagrange multipliers can be calculated quickly using a Newton-Raphson method. For example, with three bodies, the computational time is $\sim 10^{-3}$ of the total computational time. The details are given by Kajtar and Monaghan (2008). Extending the algorithm to 4 or more bodies is straightforward. 

\subsection{The rate of change of density}

The rate of change of density of the fluid particles is 
\begin{equation}
\frac{d\rho_a}{dt} = \sum_\eta m_\alpha {\bf v}_{a \eta} \cdot \nabla_a W_{a \eta},
\end{equation}
where the summation is over the labels of all the fluid, rigid boundary and skin particles. In some formulations of SPH the summation is only over the fluid particles, but a better estimate of the velocity divergence, and therefore the rate of change of the density, is obtained by including the velocity of the boundary and skin particles.  As mentioned earlier the inviscid fluid equations can be obtained from a variational principle using the continuity equation as a constraint.  As a result the pressure terms in the acceleration equation of the fluid  then involve all the particles.  The density of the boundary force particles and the skin particles is kept fixed.  In practice the changes in density are small, but they need to be correctly calculated to ensure that the pressure is estimated accurately.

The $h$ associated with any fluid particle can be obtained from $h \propto 1/\rho^{1/2}$ though we calculate it in step with the density from
\begin{equation}
\frac{dh}{dt } = -\frac{h}{2 \rho} \frac{d \rho}{dt}.
\end{equation}

\subsection{Equation of state and viscosity}\label{sec:eqnstate}

The fluid is assumed to be slightly compressible with an equation of state given by
\begin{equation}
P_a = \frac{ \rho_0 c_a^2}{7} \left (  \left(  \frac{\rho_a}{\rho_0} \right )^7 -1   \right ),
\end{equation}
where $\rho_0$ is the reference density of the fluid. To ensure the flow has a sufficiently low Mach number to approximate  a constant density fluid accurately, we determine the speed of sound by
$c_a \sim 10 V$ where $V$ is the maximum speed of the fluid relative to the bodies. In this paper we take $V = 2a \omega$ where $a$ is the semi major axis of the ellipse and $\omega$ is the frequency of the oscillation associated with the gait. The pressures of the body force particles and the skin particles are set to zero. 

The viscosity is determined by $\Pi_{ab}$ for which we choose the form (Monaghan 1997,  2005)
\begin{equation}
\Pi_{ab} = -\frac{ \alpha v_{sig} {\bf v}_{ab} \cdot {\bf r}_{ab}   }{{\bar \rho}_{ab} |{\bf r}_{ab}|}.
\end{equation}
In this expression $\alpha $ is a constant, and the notation ${\bf v}_{ab} = {\bf v}_a - {\bf v}_b$ is used. ${\bar \rho}_{ab}$ denotes the average density $\frac12(\rho_a + \rho_b)$.  We take the  signal velocity to be 
\begin{equation}
v_{sig} = \frac12( c_a + c_b)(1+\alpha) - 2\frac{ {\bf v}_{ab} \cdot {\bf r}_{ab}}{r_{ab} },
\end{equation} 
where $c_a$ is the speed of sound at particle $a$ (Monaghan 1997, although here we take $v_{sig}$ to be half used in that paper and $\alpha$ is therefore a factor 2 larger). The kinematic viscosity can be estimated by taking the continuum limit which is equivalent to letting the number of particles go to infinity while keeping the resolution length $h$ constant.  By a calculation similar to that  in Monaghan (2005) it is found for the Wendland kernel that the kinematic viscosity is \begin{equation}
\nu = \frac{1}{8} \alpha h v_{sig}.
\end{equation}
SPH calculations for shear flow agree very closely with theoretical results using this kinematic viscosity (Monaghan 2006).  Using these results we can write (2.27) in the following form
\begin{equation}
\Pi_{ab} = -\frac{ 8 \nu {\bf v}_{ab} \cdot {\bf r}_{ab}   }{{\bar \rho}_{ab} {\bar h}_{ab}|{\bf r}_{ab}|}.
\end{equation}
If desired  $\nu$ can be replaced by using the Reynolds number.
\subsection{Motion of the particles}

The position of any fluid or skin particle is found by integrating
\begin{equation}
\frac{ d{\bf r}}{dt} = {\bf v}.
\end{equation}

The motion of a boundary particle can be determined from the motion of centre of mass and the rotation about the centre of mass. Thus for particle $j$ on body $k$,
\begin{equation}
\frac{  d{\bf r}_j}{dt} = {\bf V}_k  + \Omega_k \hat{\bf z} \times ( {\bf r}_j - {\bf R}_k),
\end{equation}
where, in this two dimensional problem, the rotation is around the $z$ axis which is perpendicular to the plane of the motion.

\subsection{The gait}

A biological creature swims through a fluid by changing its shape. The oscillatory motion of a fin, for example, acts to propel the creature in a forward motion. Slight variations to the motion allow it to accelerate, decelerate, and to turn.  Fish such as eels have a gait which is similar to a wave travelling from head to tail with moderately large amplitude along the entire length. A fish such as a mackerel has a gait which is similar to travelling wave with small amplitude until roughly half way down the body when it increases sharply.  For the present two-dimensionsal, three-body swimmer considered here, the motion depends upon the angles $\varphi_1$ and $\varphi_2$. The particular specification of these angles is referred to as the `gait'. The two gaits considered here are similar to that of an eel.  

The forward gait of motion was specified with
\begin{eqnarray}
\varphi_1 &=&\theta_2(0) - \theta_1(0) +  \beta( \cos(\omega t) -1) \label{eqn:phi1}, \\
\varphi_2 &=&\theta_3(0) - \theta_2(0) + \beta \sin(\omega t) \label{eqn:phi2},
\end{eqnarray}
where for these calculations $\beta = 1$ and $\omega = 1$. We choose $\theta_1(0) = -\beta$,  $\theta_2(0) = 0$ and $\theta_3(0) = 0$. Note that this specification is identical to that of Kanso et al. (2005) and Eldredge (2007), but  the notation is different.  The turning gait is the same except  $\theta_1(0) = 0$, $\theta_2(0) = 0$ and $\theta_3(0) = -\beta$ and we take $\beta=1$ and $\omega = 1$.  With this gait   the angles $\varphi_j$ are never positive. This turning specification is the same as that of Kanso et al. (2005).

\subsection{The kernel}
In this paper we use the fourth order Wendland kernel for two dimensions (Wendland 1995). With $q=r/h$ this kernel is given by 
 \begin{equation}
 W(|{\bf r }|,h) = \frac{7}{64 \pi h^2} (1+2q)(2-q)^4,
 \end{equation} 
 when $q \le 2$ and zero otherwise.  We take the initial $h=1.5dp$.
\subsection{The initial conditions}

In the present simulations the liquid SPH particles were initially placed on a grid of squares  of side $dp$ which defines the liquid particle spacing.  Those  at least $dp$ outside  the boundary  formed by the ellipses and the skin were retained.  the boundary particles had a spacing $\Delta = dp/n$ where $n$ was typically 2. The mass of the fluid particles was $\rho_0 dp^2$, and the mass of the boundary particles was $1/n$ of this mass.  

The boundary force particles on the body were placed around each ellipse with a spacing as close as possible to $dp/n$.  Having chosen which body particles connect to the skin, the skin sections were placed on straight lines a indicated in Figure 1. The fluid particles are not initially in equilibrium with the boundary forces so we allow them, and the skin particles, to move under damping.  The rule for damping is given in the following section.

After the damping is finished, the motion starts with the initial conditions set so that the fluid and skin particles have zero velocity and the bodies have zero net angular momentum and linear momentum consistent with the time derivatives of the constraints. The details of this are given by Kajtar and Monaghan (2008). 


\subsection{The time stepping}
The time stepping is based on the second order symplectic integrator often called the Verlet integrator.  The basic equations we integrate take the following form for a liquid particle. Throughout this section, for any quantity $\bf A$, $\mathbf{A}^0$ denotes its value at the beginning of the time-step, $\mathbf{A}^{1/2}$ at the mid-point, and $\mathbf{A}^{1}$ at the end of the step. The other particles do not change their density.

\begin{eqnarray}
	\frac{d\mathbf{r}_a}{dt} &=& \mathbf{v}_a, \\
	\frac{d\mathbf{v}_a}{dt} &=& \boldsymbol{\mathcal{F}}_a, \\
	\frac{d\rho_a}{dt} &=& {\mathcal{D}}_a,
\end{eqnarray}

In the first stage of the integration, the mid-point values are calculated for  ${\bf r}_a$,  $\rho_a$ and $h_a$, the body positions and orientations $\mathbf{R}_k$,  $\theta_k$, and the relative boundary  particle positions ${\bf d}_j = {\bf r}_j - {\bf R}_k$. With $\delta t$ denoting the time step
\begin{eqnarray}
	\mathbf{r}_a^{1/2} &=& \mathbf{r}_a^0 + \tfrac{1}{2}\delta t \mathbf{v}_a^0 \\
	\rho_a^{1/2} &=& \rho_a^0 + \tfrac{1}{2}\delta t {\mathcal{D}}_a^0 \\
	h_a^{1/2} &=& h_a^0 \left(1-\frac{1}{4}\delta t {\mathcal {D}}_a^0/\rho_a^0 \right)  \\
	\mathbf{R}_k^{1/2} &=& \mathbf{R}_k^0 + \tfrac{1}{2}\delta t\mathbf{V}_k^0 \\
	\theta_k^{1/2} &=& \theta_k^0 + \tfrac{1}{2}\delta t \Omega_k^0 \\
	\mathbf{d}_j^{1/2} &=& \mathbf{d}_j^0 + \tfrac{1}{2}\delta t \Omega_k^0\mathbf{\hat{z}} \times \mathbf{d}_j^0 \\
	\mathbf{r}_j^{1/2} &=& \mathbf{d}_j^{1/2} + \mathbf{R}_k^{1/2} \end{eqnarray}
With the mid-point coordinates known,  $\boldsymbol{\mathcal{F}}_a^{1/2}$, $\mathbf{f}_j^{1/2}$, $\mathbf{F}_k^{1/2}$ and $\tau_k^{1/2}$ can be calculated.  The last two involve the Lagrange multipliers and their calculation is discussed by Kajtar and Monaghan (2008).

The  time-step for ${\bf v}$ and $\bf  r$ is then completed by 
\begin{eqnarray}
	\mathbf{v}_a^1 &=& \mathbf{v}_a^0 + \delta t \boldsymbol{\mathcal{F}}_a^{1/2}, \\
	\mathbf{r}_a^1 &=& \mathbf{r}_a^{1/2} + \tfrac{1}{2}\delta t \mathbf{v}_a^1.
\end{eqnarray}	
With ${\bf v}^1$ and ${\bf r}^1$ known $D^1_a$ can be calculated (this requires another sweep over the particles) and the step for $\rho$ and $h$ completed according to
\begin{eqnarray}	
	\rho_a^1 &=& \rho_a^{1/2} + \tfrac{1}{2}\delta t {\mathcal{D}}_a^1 ,\\
	h_a^1 &=& \frac{h_a^{1/2}}{1 + \tfrac{1}{4} \delta t \left( {\mathcal{D}}_a^1/\rho_a^1  \right)} .
\end{eqnarray}
The step for the body velocity, coordinates, angles and angular velocity is completed by 
\begin{eqnarray}
	\mathbf{V}_k^1 &=& \mathbf{V}_k^0 + \frac{\delta t}{M_k} ,
    	\left(\sum_{j\in S_k} \mathbf{f}_j^{1/2} + \mathbf{F}_k^{1/2}\right) ,\\
	\mathbf{R}_k^1 &=& \mathbf{R}_k^{1/2} + \tfrac{1}{2}\delta t \mathbf{V}_k^1, \\
	\Omega_k^1 &=& \Omega_k^0 + \frac{\delta t}{I_k} \left(\sum_{j\in S_k} \mathbf{d}_j^{1/2}\times \mathbf{f}_j^{1/2} + \tau_k^{1/2} \right) ,\\
	\theta_k^1 &=& \theta_k^{1/2} + \tfrac{1}{2}\delta t \Omega_k^1 ,
\end{eqnarray}
and the positions and normals of the  body boundary particles at the end of the step are given by
\begin{eqnarray}
\mathbf{d}_j^1 &=& \frac{\mathbf{d}_j^{1/2} + \tfrac{1}{2}\delta t \Omega_k^1\mathbf{\hat{z}} \times \mathbf{d}_j^{1/2}} {1 + (\tfrac{1}{2}\delta t \Omega_k^1)^2}, \\
\mathbf{r}_j^1 &=& \mathbf{d}_j^1 + \mathbf{R}_k^1.
\end{eqnarray}

The damping is achieved by replacing (2.45) by 
\begin{equation}
\mathbf{v}_a^1 = (\mathbf{v}_a^0 + \delta t \boldsymbol{\mathcal{F}}_a^{1/2}){\widehat D}, 
\end{equation}
where 
\begin{equation}
{\widehat D} = 1 - e^{-w}.
\end{equation}
The function $w$ is given by
\begin{equation}
w= \frac{10(n_d - n')}{n_d},
\end{equation}
where $n_d$ is the number of damping steps (typically $\sim2000$), and $n'$ is the current step. $\widehat D$ is set to 1 for $n'> n_d$.  The damping steps may seem large, but for these calculations which involve $\sim12000$ steps it is not significant.  However, it would be desirable to have more efficient damping.

The time step size, $\delta t$, is updated at the end of each time-step by
\begin{equation}
\delta t = \frac{1}{2} \text{min} \left(\frac{h_{ab}}{v_{sig}}, \frac{r_{ab}}{V}, \frac{\Delta_s}{v_s} \right),
\end{equation}
where the minimum is over all fluid, boundary and skin particles evaluated at the mid-point of the time step. The first  and last terms are CFL conditions for wave propagation in the fluid and in the skin respectively. The second term, $r_{ab}/V$, ensures that $\delta t$ is sufficiently small to follow the motion of particles very close to a boundary. 

\section{Numerical tests}
\setcounter{equation}{0}

In the absence of skin our algorithm has been tested (Kajtar and Monaghan 2008, 2010) by detailed comparison against  the results of experiments and those obtained by other authors. These include the motion of a tethered cylinder in a channel, the forced oscillation of a cylinder,  and the inviscid calculations of Kanso et al., (2005) and Melli et al., (2006) where our SPH results showed convergence to the inviscid results at Reynolds numbers of $\sim 5000$. They were also in good agreement with  the viscous calculations  of Eldredge (2006, 2007, 2008). 

In this paper we begin with a test of our model of elastic skin by following  the approach to equilibrium of fluid in a tank with an elastic skin bottom. We confirm the convergence of the calculations with finer resolution and show the the final displacement agrees with approximate theory.  We then describe the simulated motion of linked ellipses with and without skin.
\subsection{Static tank with an elastic base}

The tank had depth $D=1$ and width $L=1$, while the  fluid density was $\rho=1000kg/m^2$ (because the system is two dimensional a unit thickness in the third dimension is assumed)  giving a total fluid mass of 1000 $kg$.  The tension of the skin was  $T=2\rho g D$ and its  density $\rho_s = 1000 kg/m^2$, and thickness $\ell_s = 0.05m$. With these parameters  the speed of wave propagation along the skin is $v_s \simeq 19.8  m/s$. The Reynolds number is $Re = 50$.  We take the speed of sound to be $c= 10\sqrt{gD}$.  The fluid particles were placed on a grid of squares.  In order to determine the convergence the calculations were run for a number of different initial particle spacings in the range $dp = 1/20$ to 1/60.  The ratio of the fluid particle spacing to the boundary (and skin) particle spacing was 2. 

The skin was initially horizontal and when released the skin and fluid began damped oscillations which were followed until the skin was in equilibrium. The variation of the period of oscillation with the square of the initial particle spacing is shown in Figure 3.  The convergence is second order.  The final position of the centre of the skin can be estimated by linearizing the equations of equilibrium of an elastic skin.  Solving these equations we find that the displacement  of the skin $\eta(x) $ is given by 
\begin{equation}
\eta(x) = \frac{g \rho(D +  \ell_s)}{2T} x(x-L).
\end{equation}
At the highest resolution the SPH result for the skin displacement at $x=L/2$ is -0.068, which differs from the value -0.066  from (3.1) by 3 percent, which is satisfactory bearing in mind that  (3.1) is only approximate.

\begin{figure}[htbp]
\begin{center}
\includegraphics[width=0.8\textwidth]{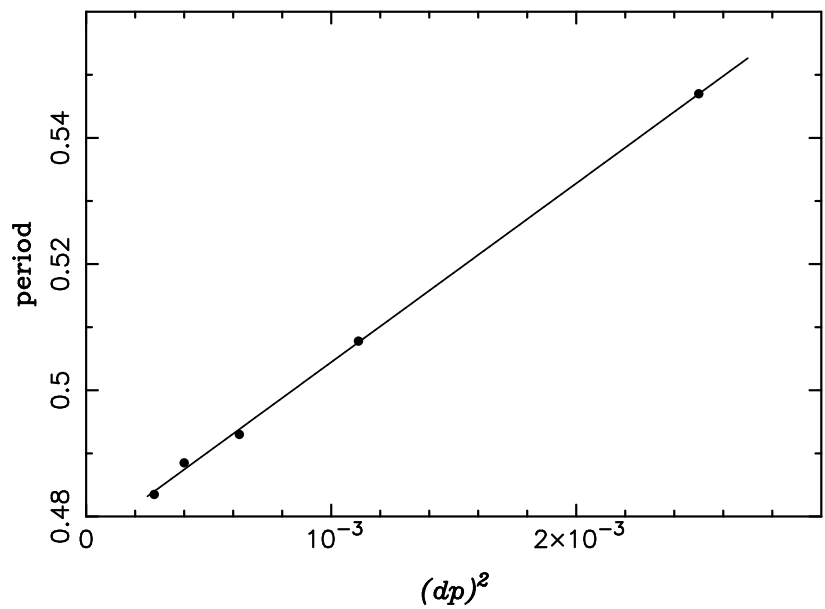}
\caption{The period of the damped oscillations of the skin forming the base of a tank of fluid. The horizontal axis is the square of the initial particle spacing $dp$.  The convergence is clearly second order in $dp$.}
\label{fig:elasticbaseerror}
\end{center}
\end{figure}

\subsection{Motion of the linked bodies}
Our aim is to compare the motion of the linked bodies, with skin and without, for both the forward and turning gaits. In the first set of tests, the linked elliptical bodies were of equal size and mass. The ellipses had semi-major axis $a=0.25$, semi-minor axis $b=0.2a$, and distance between the tip of the ellipse and the pivot $c=0.2a$. The second set of tests considered unequal sized bodies, but with the same total mass as for the first set. The body length parameters were $a_1 = 0.25$, $a_2 = a_3 = 0.5a_1$, $b_1 = 0.4a_1$, $b_2 = 0.5a_2$, $b_3 = 0.3a_3$, and for all bodies $c=0.2a_1$. In all cases, the densities of the bodies were the same as the fluid, $\rho = 1000$.

The Reynolds number $\Re$ is defined by using the characteristic velocity $V = 2a\omega$ and the characteristic length $L=2a$, so that
\begin{equation}
\Re = \frac{4a^2\omega}{\nu}.
\end{equation}
In the present simulations $\Re = 200$. The speed of sound was $c_s = 20 a \omega$, and the boundaries of the ellipses were defined by boundary particles with spacing as close to $dp/2$ as possible. The motion takes place in a domain with periodic rectangular cells. Based on the convergence studies of Kajtar and Monaghan (2010), the initial particle spacing was chosen to be $dp = 1/60$. The fluid spans from $x_\mathrm{min} = 0$ to $x_\mathrm{max}$ along the horizontal axis, and from $y_\mathrm{min} = 0$ to $y_\mathrm{max}$ in the vertical axis. For the forward gait, the domain was of size $x_\mathrm{max} \times y_\mathrm{max} = 6 \times 4.5$, and the initial coordinates of the centre of mass of the middle body were $(X_2,Y_2) = (0.4x_\mathrm{max}, 0.6y_\mathrm{max})$. For the turning gait, the domain was $x_\mathrm{max} \times y_\mathrm{max} = 5 \times 5$, and the initial coordinates were $(X_2,Y_2) = (0.4x_\mathrm{max}, 0.5y_\mathrm{max})$.

\begin{table}
\begin{center}
\begin{tabular}{|c|c|c|c|}
\hline 
Body & 1 & 2 & 3 \\ 
\hline
$\Omega$ & $-2.771\times 10^{-1}$ & $-2.771\times 10^{-1}$ & $7.229\times 10^{-1}$ \\ 
\hline
$\dot X$ & $-4.664\times 10^{-2}$ & $2.332\times 10^{-2}$ & $2.332\times 10^{-2}$  \\ 
\hline
$\dot Y$ & $-4.079\times 10^{-2}$ & $8.726\times 10^{-2}$ & $-4.647\times 10^{-2}$ \\ 
\hline
\end{tabular}
\caption{The initial velocities for the forward gait}
\label{tab:fgvel}
\end{center}
\end{table}

\begin{table}
\begin{center}
\begin{tabular}{|c|c|c|c|}
\hline 
Body & 1 & 2 & 3 \\ 
\hline
$\Omega$ & $-2.143\times 10^{-1}$ & $-2.143\times 10^{-1}$ & $7.857\times 10^{-1}$ \\ 
\hline
$\dot X$ & $6.612\times 10^{-2}$ & $6.612\times 10^{-2}$ & $-1.322\times 10^{-1}$  \\ 
\hline
$\dot Y$ & $-6.469\times 10^{-2}$ & $6.388\times 10^{-2}$ & $-8.063\times 10^{-4}$ \\ 
\hline
\end{tabular}
\caption{The initial velocities for the turning gait}
\label{tab:tgvel}
\end{center}
\end{table}

\begin{figure}[htbp]
\begin{center}
\includegraphics[width=0.7\textwidth]{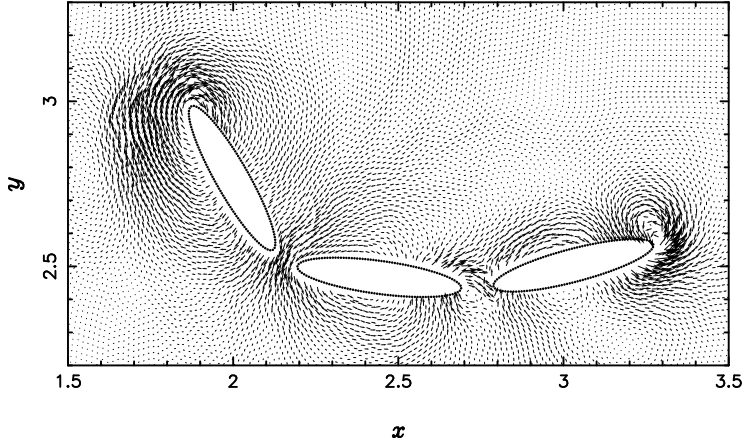}
\caption{The velocity field for the case of forward motion in the absence of skin and equal sized ellipses.}
\label{fig:bodyandnoskin}
\end{center}
\end{figure}

The skin segments were attached to the ellipses at the mid-point between where the major and minor axes intercept the ellipse. In other words, the last skin particle belonging to a skin segment was connected to the body particle 1/8 of the distance around the ellipse from the major axis. Since the skin thickness was chosen as $\ell_s = 0.05$, the skin segments had a significant mass. For example, with equal sized bodies, the total skin mass was approximately 50\% of the total mass of the three bodies. In order to account for this added mass, the body masses were reduced so that the total mass of the swimmer with skin or without was the same. For the cases where the body masses were unequal, their masses were reduced such that the body densities remained equal. The initial linear velocities and angular momentum are given in Table 2.

The effect of the skin on the velocity field is shown by Figures 4 and 5.  The flow around the tips of the end points is very similar in each case, but Figure 4 shows a flow between the gaps which influences the flow along the body. 

\begin{figure}[htbp]
\begin{center}
\includegraphics[width=0.7\textwidth]{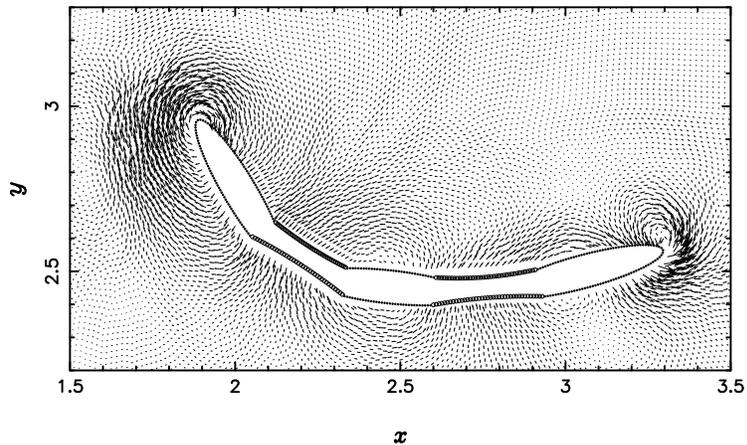}
\caption{The velocity field for the case of forward motion with skin and equal sized ellipses. The skin is shown by a dark line which can be best viewed under magnification as provided by Adobe.}
\label{fig:vel-bodyandskin}
\end{center}
\end{figure}

Table \ref{tab:fgres} shows the results for the forward gait, with skin and without, and for equal and unequal body sizes. Table \ref{tab:tgres} shows the results for the turning gait. For the forward gait, the distance travelled by the swimmer in three periods (or $t=6\pi$) was taken as the length between the initial and final positions of the centre of mass of the middle body, $\mathbf{R}_2$. The angle of motion (in radians) was taken as the angle of this length relative to the $x$ axis. For the turning gait, the amount of rotation was taken as the angle of the middle body relative to the $x$ axis after three periods (or $t=6\pi$). Following Kajtar and Monaghan (2010), the average power $\mathcal{P}$ expended by the swimmer was computed by numerically integrating the following expression
\begin{equation}
{\mathcal P} = \frac{1}{6\pi} \int _{0}^{6\pi}\left( \lambda_\theta^{(1)} ( \Omega_2 - \Omega_1 ) + \lambda_\theta^{(2)} ( \Omega_3 - \Omega_2 ) \right) dt.
\end{equation}

\begin{figure}[htbp]
\begin{center}
\includegraphics[width=0.8\textwidth]{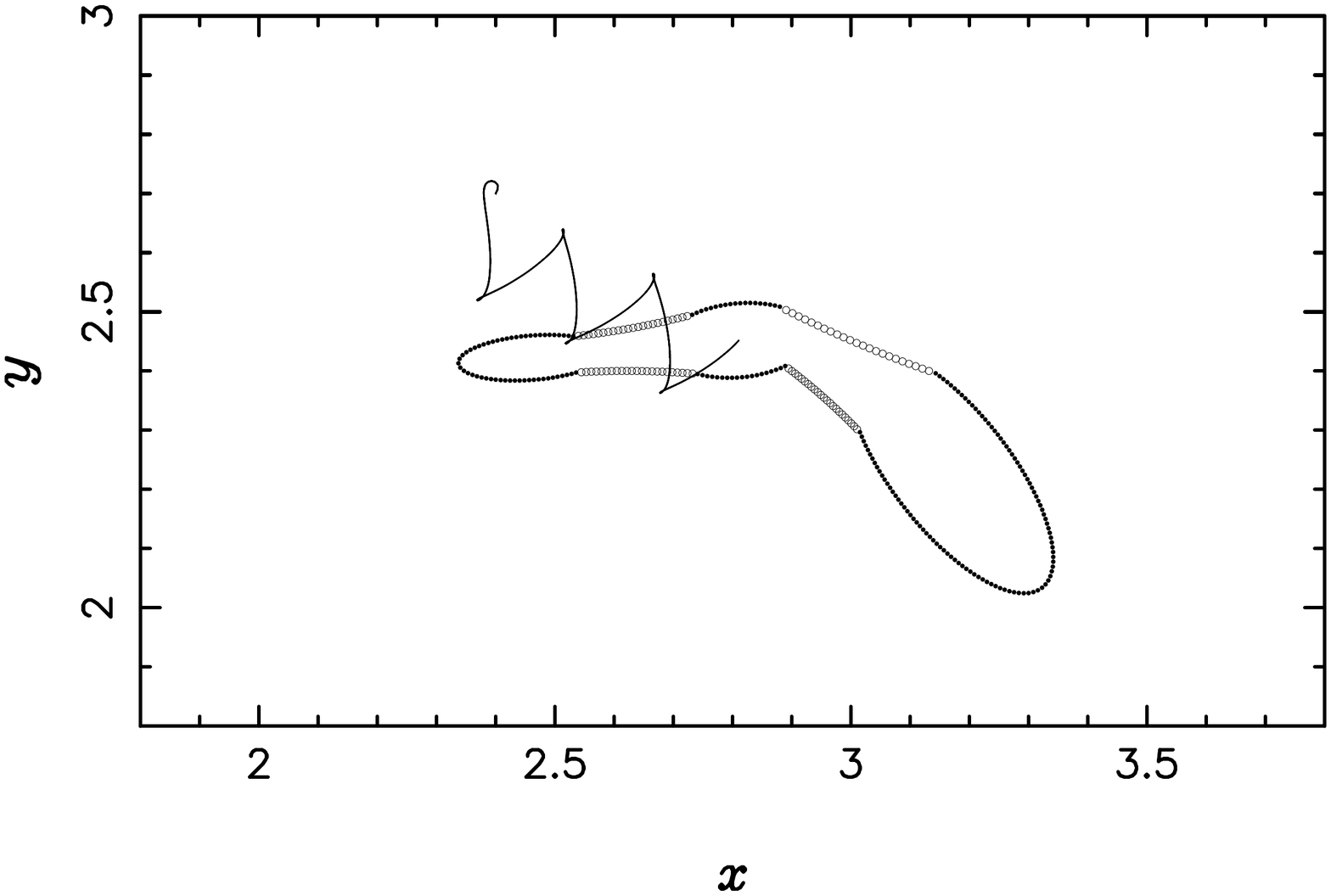}
\caption{The zig-zag path of the centre of mass of the central body in the case where there is skin and the ellipses are of different size. The position of the bodies is shown for the last point of the zig-zag.}
\end{center}
\end{figure}
\begin{table}
\begin{center}
\begin{tabular}{|c|c|c|c|c|}
\hline 
body sizes & skin & distance & $\mathcal{P}$ & angle \\ 
\hline
\multirow{2}{*}{equal} & no  & 0.7038 & 83.65 & -0.289 \\ 
& yes & 0.7925 & 74.63 & -0.299 \\ 
\hline
\multirow{2}{*}{unequal} & no  & 0.3548 & 24.97 & -0.525 \\ 
& yes & 0.4799 & 21.54 & -0.541 \\ 
\hline
\end{tabular}
\caption{Distance travelled by the swimmer using the forward gait, in time $t=6\pi$, with skin and without, and for equal and unequal body sizes. The power expended and angle of motion are also given.}
\label{tab:fgres}
\end{center}
\end{table}
\begin{figure}[htbp]
\begin{center}
\includegraphics[width=0.8\textwidth]{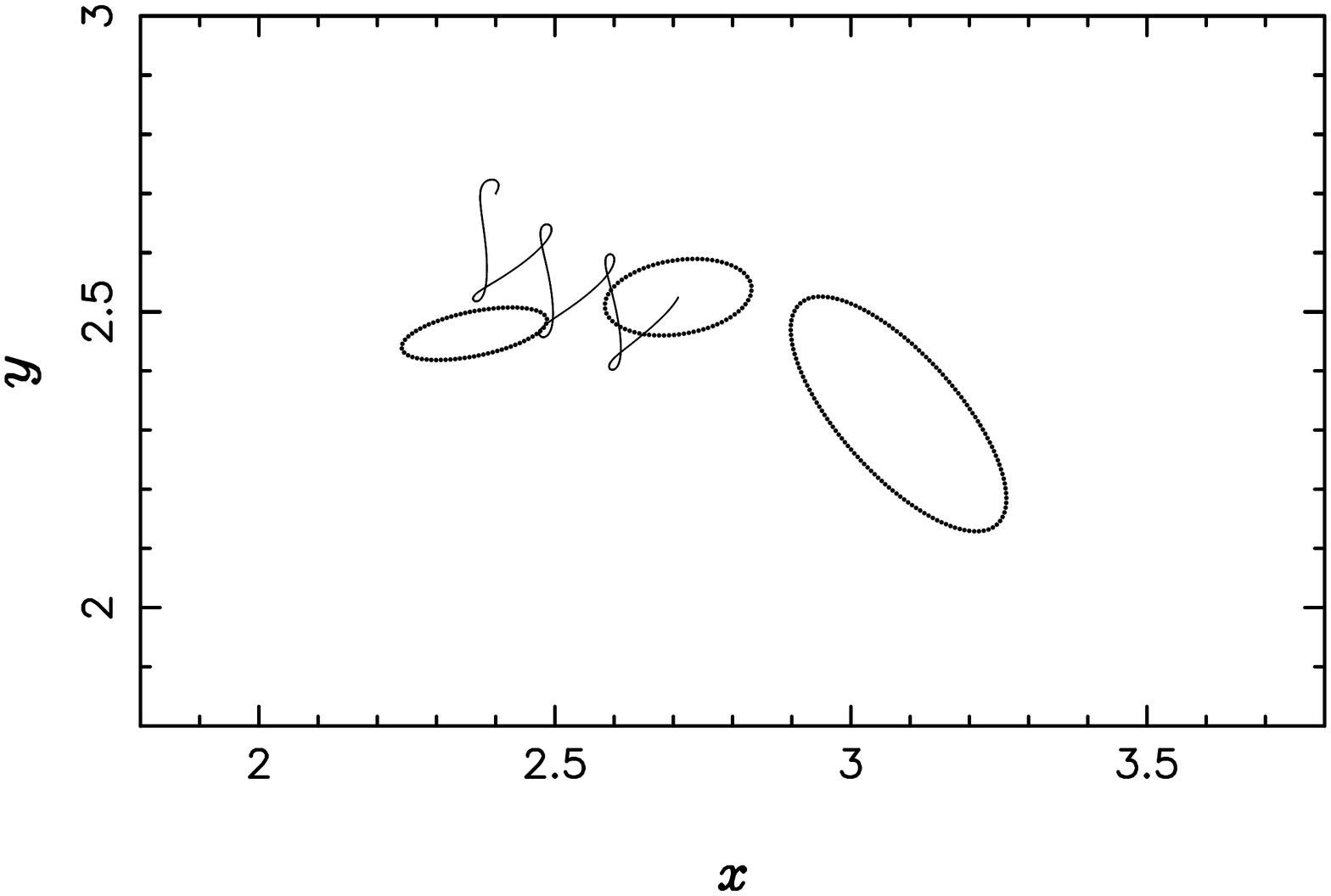}
\caption{The velocity field for the case of forward motion  in the absence of skin. The position of the bodies is shown for the last point of the zig-zag.}
\label{fig:xy-bodyandskin}
\end{center}
\end{figure}
Figure 6 shows the zig-zag path of the centre of mass of the central body in the case of skin. Comparison with Figure 7 shows that the presence of skin results in a sharper loop at the extremes of the zig-zag, and the overall path is at a steeper angle to the horizontal.

\begin{table}
\begin{center}
\begin{tabular}{|c|c|c|c|c|}
\hline 
body sizes & skin & rotation & $\mathcal{P}$\\
\hline
\multirow{2}{*}{equal} & no & 3.7532 & 142.52 \\ 
& yes & 4.0487 & 135.60 \\ 
\hline
\multirow{2}{*}{unequal} & no & 4.0653 & 56.58 \\ 
& yes & 3.4988 & 49.48 \\ 
\hline
\end{tabular}
\caption{Amount of rotation (measured in radians) using the turning gait, in time $t=6\pi$. The power expended is also given.}
\label{tab:tgres}
\end{center}
\end{table}

For the forward gait, it is clear that the inclusion of skin improves the efficiency of the swimming. For equal body masses, the motion is improved by almost 13\%, and for unequal masses by $\sim 39\%$.  We relate these improvements to the skin preventing fluid passing between the bodies. Although the swimmer with unequal body masses travels approximately half the distance in the same period of time, the power expended is reduced by a factor of $\sim 3$. In this study, the gait, $\beta$, and $\omega$ were identical. Kajtar and Monaghan (2010) showed that the efficiency of the swimming depends upon these parameters. Undoubtedly the efficiency of the swimmer with unequal body masses would also improve for a different set of parameters. Swimming with skin or without only changes the angle of motion by $\sim 3\%$, but the mass distribution has a much larger effect.

For the turning gait with equal body masses, the middle body rotates $\sim 8\%$ further when the skin is included. As with the forward gait, the power expended is slightly reduced. Without skin, the swimmer with unequal body masses turns $\sim 8\%$ further. However, with the skin included, it rotates less.

\section{Conclusions}     
In this paper we have described a Lagrangian based SPH algorithm which allows us to simulate a system consisting of fluid, rigid body and elastic skin.  The imposed change of angle between the bodies is included by using Lagrange multipliers.  By basing the time stepping on a symplectic integrator  the linear momentum  and angular momentum of the system are conserved  to high accuracy; for the former the errors are due to round off  and for the latter they are dominated by the effect of the periodic boundary conditions which do not conserve angular momentum for a particle system. The algorithm allows us to treat both straight line and turning motions and bodies of arbitrary shape and relative size.   We have applied the algorithm to three linked ellipses which may be of different size, and shown that the presence of the skin affects the speed, power output and turning capacity of the bodies.  There is no difficulty in extending the code to deal with more bodies.  

Within the limits imposed by the algorithm being two dimensional our results suggest many interesting applications including the study of bodies swimming through a free surface, or in a stratified medium, and the analysis of the hunting gaits of predators and the escape gaits of their prey.

\section{Appendix}
\setcounter{equation}{0}
In this appendix we  write out the Lagrangian for the equations of motion in the absence of viscous forces, and show how the variational principle with the continuity equation as a constraint gives the pressure forces.  

\subsection{The Lagrangian}

The Lagrangian ${\mathcal {L}}$ consists of the following terms

\begin{equation}
{\mathcal {L}} = {\mathcal {L}}(fluid) + {\mathcal {L}}(bodies) +{\mathcal {L}}(skin) + {\mathcal {L}}(int),
\end{equation}
where the first three terms have the form
\begin{equation}
{\mathcal {L}}(fluid) = \sum_b m_b \left(  \frac12 v_b^2 - u(\rho)  \right),
\end{equation}
\begin{equation}
{\mathcal {L}}(bodies) = \sum_{k=1}^{N_b}  \left( \frac12   M_k v_b^2  + \frac12 I_k \Omega_k^2)  \right),
\end{equation}
and
\begin{equation}
{\mathcal {L}}(skin) = \sum_\sigma \left ( \frac12 m_\sigma v_\sigma^2 - \frac12 \sum_{\sigma'} \kappa ({\bf r}_{\sigma \sigma'} )^2 \right),
\end{equation}
The fourth term involves the interactions between the particles.  The first step is to note that the terms  $ {\bf r}_{aj}f(|{\bf r}_{aj}| )$ can be written in terms of the gradient of a potential according to 
\begin{equation}
 {\bf r}_{aj}f(|{\bf r}_{aj}| ) = - \frac{\partial  \Phi(| {\bf r}_{aj} |}{\partial {\bf r}_a},
\end{equation}
where  $\Phi(| {\bf r}_{aj} |)$ is a potential energy which we will denote by $\Phi_{a j}$.  The interaction part of the Lagrangian is then given by
\begin{equation}
{\mathcal {L}}(int) = - \sum_b \sum_{k=1}^{N_b} \sum_{j \in B(k) }m_bm_j \Phi_{bj}- \sum_b \sum_{k=1}^{N_s} \sum_{j\sigma \in S(k) }m_b m_\sigma \Phi_{b\sigma} -\sum_\sigma \sum_j\frac12 \kappa {\bf r}_{\sigma j}^2,
\end{equation}
where in the last term the values of $\sigma$ and $j$ are those for connected pairs.  Following the usual rules the inviscid equations of motion can be worked out.  Because the Lagrangian is invariant to translations and rotations  of the coordinate system the linear and angular momentum are conserved.  A discrete version of Kelvin's circulation theorem can also be deduced (Monaghan 2005).  In order to work out Lagrange's equations for the rigid bodies it is necessary to relate the change in position of the centres of mass of the bodies and their angles $\theta$ to changes in the positions of the body particles.  To do this we note from (2.32) that for body force particle $j$ on body $k$ 
\begin{equation}
  \delta {\bf r}_j =  \delta {\bf R}_k  + ( \delta \theta_k) \hat{\bf z} \times ( {\bf r}_j - {\bf R}_k).
\end{equation}

Because the continuity equation must be satisfied it acts as  a constraint when using the least action principle.  We consider this next. 
\subsection{Least action and the Continuity equation}
In this section we consider a purely fluid dynamical problem with the Lagrangian 
\begin{equation}
{\mathcal {L}}= \sum_b m_b \left(  \frac12 v_b^2 - u(\rho)  \right),
\end{equation}
which is to be substituted in the least action principle and varied with the continuity equation
\begin{equation}
\frac{d\rho_b}{dt} = \sum_\eta m_\eta ({\bf v}_{b} - {\bf v}_\eta )\cdot \nabla_b W_{b \eta},
\end{equation}
acting as a constraint.  The summation over $\eta$ denotes a summation over all the particles.
The variational principle of least action results in the equation of motion of particle $a$
\begin{equation}
\frac{d}{dt } \left ( \frac{\partial {\mathcal {L}}}{\partial {\bf v}_a}  \right ) = \frac{\delta {\mathcal {L}}}{\delta{\bf r}_a },
\end{equation}
where $\delta$ denotes a Lagrangian change.  We can write 
\begin{equation}
\frac{\delta {\mathcal {L}}}{\delta{\bf r}_a } = -\sum_b m_b \frac{P_b}{\rho_b^2} \left (  \frac{\delta \rho_b}{\delta{\bf r}_a }\right ),
\end{equation} 
and note from the continuity equation that
\begin{equation}
\delta \rho_b = \sum_\eta ( \delta {\bf r}_b - \delta {\bf r}_\eta) \cdot \nabla_b W_{b \eta}.
\end{equation}
From the previous equation we deduce that
\begin{equation}
 \frac{\delta \rho_b}{\delta{\bf r}_a } = \sum_\eta (\delta_{ba} - \delta_{\eta a}) \nabla_b W_{b \eta},
\end{equation}
where $\delta_{ab}$  is a Kronecker delta which is 1 if $a=b$ and zero otherwise. Substitution of (4.13) into (4.11) then gives
\begin{equation}
\frac{\delta{\mathcal {L}}}{\delta{\bf r}_a } = -\frac{m_a P_a}{\rho_a^2} \sum_\eta m_\eta \nabla_a W_{a \eta} + \sum_b \frac{m_b P_b}{\rho_b^2}  \nabla_bW_{ab}.
\end{equation}
The first term is summed over all the particles whereas the second term is summed only over the fluid particles.  However, the pressure assigned to the boundary particles and the skin particles is zero so we can extend the second summation over all the particles. Noting that $\nabla_a W_{ab} = -\nabla_b W_{ab}$ we can finally write 

\begin{equation}
\frac{\delta{\mathcal {L}}}{\delta{\bf r}_a } = - \sum_\eta m_\eta \left(  \frac{ P_a}{\rho_a^2} +  \frac{ P_\eta}{\rho_\eta^2}  \right) \nabla_a W_{a \eta} .
\end{equation}
This gives the pressure terms on the right hand side of (2.1) where it has been split into separate contributions from the fluid particles, the body particles, and the skin particles.  


\end{document}